\documentstyle[12pt]{article}
\font\bbexii=cmmib10 scaled\magstep1
\font\bbex=cmmib10
\font\bbevii=cmmib7

\newfam\bbefam
\textfont\bbefam=\bbexii
\scriptfont\bbefam=\bbex
\scriptscriptfont\bbefam=\bbevii
\mathchardef\BigEchar="710F
\begin{document}
\begin{titlepage}
\vspace{2.5cm}
\baselineskip 16pt
\begin{center}
\vspace{1.5cm}
\rightline{cond-mat/9805165}
\vspace{1.5cm}
\large\bf {Configuration Space}\\
\large\bf {for Random Walk Dynamics$^1$}\\
\vspace{1.5cm}
\large{
Bernd A. Berg$^{2,3,4}$ and Ulrich H.E. Hansmann$^{5,6}$
}
\end{center}
\vspace{2cm}
\begin{center}
{\bf Abstract}
\end{center}

Applied to statistical physics models, the random cost algorithm
enforces a Random Walk (RW) in energy (or possibly other 
thermodynamic quantities). The dynamics of this procedure 
is distinct from fixed weight updates. The probability for a 
configuration to be sampled depends on a number of unusual
quantities, which are explained in this paper. This has been
overlooked in recent literature, where the method is advertised
for the calculation of canonical expectation values. We illustrate
these points for the $2d$ Ising model. In addition, we prove a
previously conjectured equation which relates microcanonical 
expectation values to the spectral density.

\medskip

\vfill

\footnotetext[1]{{This research was partially funded by the
Department of Energy under contract DE-FG05-87ER40319.}}
\footnotetext[2]{{Department of Physics, The Florida State University,
                      Tallahassee, FL~32306, USA. }}
\footnotetext[3]{{Supercomputer Computations Research Institute,
                      Tallahassee, FL~32306, USA.}}
\footnotetext[4]{{E-mail: berg@hep.fsu.edu}}
\footnotetext[5]{Department of Physics, Michigan Technological University,
                 Houghton, MI~49931, USA.}
\footnotetext[6]{E-Mail: hansmann@mtu.edu}

\end{titlepage}

\baselineskip 24pt

The performance of a numerical simulation depends on the chosen weight 
factors. Hence, some attempt should be made to optimize them
for the problem at hand. The weight function of canonical MC
simulations is $\exp (-\beta E)$, where $E$ is the energy of the 
configuration to be updated and $\beta$ is the inverse temperature in 
natural units. The Metropolis algorithm and other methods generate 
canonical configurations ({\it i.e.} the Gibbs ensemble) through a 
Markov process. It has been expert wisdom~\cite{ToVa77} for quite a 
while and became widely recognized in recent years
that MC simulations with a-priori unknown weight factors are also 
feasible and deserve to be considered, for a concise recent review
see~\cite{Be97}. For instance, weighting (in a certain energy range) 
with the inverse spectral density $1/n(E)$ has turned out to be of 
practical importance. Examples are calculations of interfacial tensions
for first order phase transitions, where improvements of many orders of 
magnitude were obtained. Instead of the energy other thermodynamic 
variables can be considered as well, {\it e.g.}~\cite{all}. To be 
definite, we focus on the energy.
\medskip

MC simulations with a-priori unknown weight factors require an additional
step, not encountered in canonical simulations: A working estimate of the 
weight factors needs to be obtained first. Quite efficient recursive methods
have been developed for this purpose~\cite{Be98}. Still, the question 
suggests itself whether one can possibly bypass the first step and 
develop methods which sample broad energy distributions right away.
Indeed, it is possible to design updates such that a RW in some 
cost function is generated~\cite{Be93} and the energy of a statistical 
physics model can be chosen as cost function. Unfortunately (as already
noted in~\cite{Be93}) the connection to the desired canonical expectation
values is apparently lost. Nevertheless, it appears to be worthwhile
to investigate properties of the thus generated configurations. In 
particular, some details are subtle and, besides the origin of the 
method, ignored in recent literature~\cite{HeOl98}. 
\medskip

We consider generalized Ising models in $d$ dimensions, described by the 
energy function
\begin{equation}
E = - \sum_{ij} J_{ij}\, s_i s_j
\end{equation}
where the sum is over nearest neighbors and the exchange coupling
constants $J_{ij}$ as well as the spins $s_i,\,s_j$ take the values
$\pm1$. The Ising ferromagnet is obtained with $J_{ij} \equiv 1$.
Other special cases are the Ising anti-ferromagnet, frustrated Ising
models and spin glasses. We consider a configuration of $N$ spins and
choose periodic boundary conditions. Under flip of a single spin the 
energy can change by the following increments
\begin{equation} \label{groups}
\triangle E_i = 4\, i ~~{\rm with}~~ 
i=-d,\,\dots\,,-1,0,\,1,\,\dots\, ,d\, .
\end{equation}
\bigskip 
In the following we use $i$ to label Flip Groups (FGs) of spins.
We define now numbers 
\begin{equation} \label{FGM}
N_i ~~{\rm with}~~ i=-d,\,-d+1,\,\dots\,d-1,\,d ~~{\rm and}~~
\sum_{i=-d}^d N_i = N
\end{equation}
\bigskip
to partition the configuration of $N$ spins with respect to the FGs.  
Namely,  $N_i$  denotes the number of spins which, when flipped, 
change the energy by $\triangle E=4\,i$. 
In the following $N_i$ is referred to as Flip Group Magnitude (FGM).
The random cost algorithm~\cite{Be93} achieves a RW in energy by 
flipping spins with suitable probabilities related to FGs. Let $P_i$ be 
the probability for picking and flipping a spin in the FG labeled 
by~$i$. A RW in $E$ is obtained whenever the equation
\begin{equation} \label{RW}
\sum_{i=1}^d\, i\, \left( P_i - P_{-i} \right) = 0
\end{equation}
holds, because the expectation value of energy changes 
$\overline{\triangle E}$ becomes then zero.
In should be noted that $P_0$ does not enter this equation and
can be chosen at will. Besides, equation~(\ref{RW}) does not fix the other
probabilities either, but allows considerable freedom concerning their
further design. Before we come to this, let us note that $N_i$ has to be
greater than zero for at least one $i\ge 1$ and one $i\le -1$. Otherwise, 
a RW can no longer be achieved. This latter difficulty happens in a local 
minimum or maximum of the system and, by whatever additional rule, one or 
more spins have to be flipped before the RW simulation can continue. In 
the following we assume that the noted $N_i>0$ condition is fulfilled.
\medskip

Solutions to (\ref{RW}) are easily found, the following is given 
in~\cite{Be93}. We define
\begin{equation} \label{DEAV} \triangle E^{\,+} = {1\over N^+} 
 \sum_{i=1}^d N_i\,\triangle E_i ~~{\rm and}~~ \triangle E^{\,-} 
 = {1\over N^-} \sum_{i=1}^d N_{-i}\, \triangle E_{-i}
\end{equation}
where
\begin{equation} \label{NPM}
 N^+ = \sum_{i=1}^d N_i ~~{\rm and}~~ N^- = \sum_{i=1}^d N_{-i}\ .
\end{equation}
In the same way, we define
\begin{equation} \label{PPM}
 P^+ = \sum_{i=1}^d P_i ~~{\rm and}~~ P^- = \sum_{i=1}^d P_{-i}\ .
\end{equation}
{\it I.e.} $P^{\,+}$ is the probability to pick any of the spins from
the $i\ge 1$ FGs and $P^{\,-}$ is the probability to pick any of the spins 
from the $i\le -1$ FGs. Finally, assume that within those FGs the spins 
are picked with uniform probability, with $p^+=P^{\,+}/N^+$ for $i\ge 1$ and
with $p^-=P^{\,-}/N^-$ for $i\le -1$. The RW equation~(\ref{RW}) is then 
implied by the condition
\begin{equation} \label{RC}
 -P^{\,-}\, \triangle E^{\, -} = P^{\,+}\, \triangle E^{\,+}\ .
\end{equation}
Choosing an arbitrary probability $P_0$, the probabilities $P^{\,+}$ and 
$P^{\,-}$ follow immediately from this equation and the normalization 
condition 
$ P_0 + P^{\, +} + P^{\, -} = 1 $.
Another way~\cite{HeOl98} to implement (\ref{RW}) is to choose a spin 
at random and to reject the flip with the appropriate probability, then
counting the configuration at hand again. Here we stay with~(\ref{RC}). 
\medskip

It follows from equations (\ref{FGM}) and (\ref{RW}) that every such
algorithm samples with weights which depend on the FGMs
\begin{equation} \label{weight}
 w = w\, ( N_{-d},\, N_{-d+1},\, \dots\,, N_{d-1},\, N_d )\ .
\end{equation}
For configurations at a fixed energy $E$ the implication of this equation
is that the RW algorithm weights them differently depending on the FGM
partition, whereas canonically all these configurations have the same weight.
In the following we illustrate this point for the $2d$ Ising ferromagnet.
\medskip

We have performed canonical as well as RW simulations for $2d$ Ising
models on $N=L^2$ lattices with periodic boundary conditions. For the
RW updating we used $P_0=0.2$ and did a random flip, once the energy
minimum was reached. At large energy we imposed a cut-off~\cite{cut} 
at $E=0$, by replacing RW with random (canonical $\beta=1/(k_BT)=0$) 
updating for $E>0$.
To avoid getting lost in a flood of data, we focus on a single energy.
After gaining some experience $E/N=-1$ with canonical simulations at
$\beta=0.38$ turned out to be a reasonable choice
($\overline{E}(\beta=0.38)/N \approx -1$). This value is in the 
disordered phase for $\beta$ below the Curie point at  $\beta_c =
0.5\ln (1+\sqrt{2}) = 0.4406\dots$~. This correspond to a configuration
space region far away from the energy minimum $E/N=-2$ or the upper energy
bound $E/N=0$ imposed on the RW simulation. The lattice sizes used are 
$L=4, 10, 20, 40$ and 80. For each case we generated a statistics of
$20\times 100,000$ sweeps through the lattice and calculated error
bars with respect to twenty bins. For $L=4$ we also obtained exact results 
by simply counting through all $2^{16}$ configurations and convinced 
ourselves that the canonical simulation agrees (within very small 
statistical errors) with these exact results, whereas the RW simulation
shows already considerable deviations. 
\medskip

Let us focus on the microcanonical average values $\overline{N}_i/N$. For 
the $L=80$ lattice table~1 collects results from the canonical as well as 
from the RW simulation. Although over-all small, in case of 
$\overline{N}_{-2}/N$ the discrepancy is about 
a factor of two and for all FGs the difference between the 
canonical and the RW values clearly exceed the error bars. In the
average most spins, about 39\%, are found in the FG with number $i=2$. Hence, 
we have the best statistics for this FG and choose it to demonstrate a 
few more details. Although the discrepancy between the $\overline{N}_2/N$
values of the table appears quite small, there are considerable 
differences when we look at the distribution of $N_2$. To correct for
the expected (non-critical) finite size behavior, we define the quantity
\begin{equation} \label{n}
 n = L^{-1}\ ( \overline{N_2} - N_2 )\ .
\end{equation}
Figure~1 shows the canonical and the RW histograms $h_2(n)$, all
normalized to 
$$ L^{-1}\,\sum_n h_2(n)=1\ .$$
Error bars are negligible on the
scale of this figure. The canonical histograms for the
different lattice sizes collapse nicely into one curve, whereas the finite 
size behavior of the RW histograms fails to reproduce this behavior. The RW 
histograms are far too broad and the peak height decreases with lattice 
size, implying that the discrepancy to canonical simulation increases with
lattice size. The $L=4$ data do not fit into the scale of this figure. As 
there are only four non-zero entries, we collect them in table~2.
\medskip

By taking averages, the dependence on the FGMs appears to be washed out.
This is obvious for the average $\overline{N}_i/N$ values reported in 
table~1 and claimed~\cite{HeOl98} to be true for thermodynamic quantities 
like the energy and the specific heat. The latter quantities were calculated
from the spectral density $g(E)$ which, in turn, was determined from 
the conjectured equations
\begin{equation} \label{spectral}
\overline{N}_i (E)\, g(E) = 
\overline{N}_{-i} (E+\triangle E_i)\, g(E+\triangle E_i)\,.
\end{equation}
In appendix~A we give a mathematical proof of these equations. From this
it follows that the $\overline{N}_i(E)$ are microcanonical averages and 
{\it not} averages accumulated during the RW, as stated in~\cite{HeOl98}.
Only canonical, microcanonical or other simulations which give equal weights
to distinct configurations at the same energy will converge towards the 
correct $\overline{N}_i(E)$ values for this equation. 
A rigorous calculation of the spectral density $g(E)$ from RW data is not 
possible, because the RW weights depend on the FGMs~(\ref{FGM})~\cite{1d}. 
\medskip

For the $L=80$ lattice table~1 lists RW and canonical expectation values
for the $\overline{N}_i(E)$ at $E/N=-1$. As the differences are not too
large, it is plausible that some RW and canonical results can be in 
qualitative agreement with one another. 
However, it is obvious that the dependence of the RW configurations 
weights~(\ref{weight}) on the FGM partition $\{N_i\}$ enter the Markov 
process. Even small deviations from the canonical weights may amplify, 
because 
they enter multiplicatively through each transition step. If the RW method
is nevertheless used to estimate canonical expectation values, 
uncontrolled errors result with no guarantee that they will be negligible 
when it really matters (Murphy's law). 
\medskip

A simulation is normally already subject to finite size and other 
difficult to control approximations. Certainly, one would not like to 
build a large scale numerical investigation on a method which introduces 
an additional bias. The question arises, whether the weight 
dependence~(\ref{weight}) could eventually be controlled rigorously. Due 
to the large number of partitions of the total number of spins $N$ into 
FGMs~(\ref{FGM}) the prospects for this do not look particularly good,
but it may be worthwhile trying. Finally, we like to emphasize that the
RW approach remains a competitive method for the purpose it was 
originally~\cite{Be93} designated for, namely to find good energy
minima for optimization problems and systems with conflicting constraints.
\hfill\break 

\section*{Appendix A}

Here we prove equation (\ref{spectral}) for the $d$-dimensional 
generalized Ising mode on a lattice with periodic boundary 
condition. Let $K$ denote a spin 
configuration at energy $E$. By definition
$$ \overline{N}_i (E) = {1\over g(E)}\, \sum_K N_i (K) $$
holds. We introduce $E'=E+\triangle E_i$ and label spin configurations
at energy $E'$ by $K'$. Then
$$ \overline{N}_i (E') = {1\over g(E')}\, \sum_{K'} N_i (K') $$
and equation (\ref{spectral}) becomes equivalent to
\begin{equation} \label{Uli}
\sum_K N_i (K) = \sum_{K'} N_{-i} (K')\ .
\end{equation}
Equation (\ref{Uli}) is shown as follows. We label spins in a
fixed configuration by $s_n(K)$ where $n=1,\dots,N$ and $N$ is the
(fixed) total number of spins. In this way each single spin is
uniquely identified. The same is true for spins $s_m(K')$ in
the configurations at energy $E'$. Assume now, spin $s_{n_1}(K_1)$ 
is in flip group $i$ and we flip it. It becomes a spin 
$s_{m_1}(K'_1)$ in the flip group $-i$ at energy $E'$. No other spin, 
say $s_{n_2}(K_2)$ with at least $n_2\ne n_1$ or $K_1\ne K_2$, will
be mapped on $s_{m_1}(K'_1)$. The reason is: we can flip the spin
$s_{m_1}(K'_1)$ back and it will map precisely onto the orginal
configuration and spin, {\it i.e.} become $s_{n_1}(K_1)$. The same 
argument applies when we flip an arbitrary spin from flip group $-i$ 
at energy $E'$. Together this proves: spins in the flip groups with
magnitudes $N_i(K)$ and $N_{-i}(K')$ are in on-to-one correspondence 
and, hence, equation (\ref{Uli}) is true.
\hfill\break

\hfill\break 

\section*{Tables and Figure Captions}

\begin{table}[ht]
\centering
\begin{tabular}{||c|c|c|c|c|c||}                                        \hline
$i$& $-2$        & $-1$        & $0$         & $1$         & $2$         \\ \hline
CS &0.018853 (03)&0.072752 (04)&0.187630 (07)&0.331070 (11)&0.389694 (09)\\ \hline
RW &0.034282 (26)&0.057936 (19)&0.169412 (57)&0.350240 (43)&0.388130 (55)\\ \hline
 
\end{tabular}
\caption{Average Flip Group Magnitudes $\overline{N}_i/N$, $i=-2,\,\dots\,,2$ 
as obtained for the Canonical Simulation (CS) versus the Random Walk (RW) 
simulation on a $80\times 80$ lattice. Error bars are given in the parenthesis 
and apply to the last two digits.}
\end{table}

\begin{table}[ht]
\centering
\begin{tabular}{||c|c|c|c|c||}                                   \hline
$n$  & -1.5276     & -0.5276     & -0.0276     &  0.4724      \\ \hline
$N_2$&  0          &  4          &   6         &  8           \\ \hline
Exact&  32/424 =   & 192/424 =   & 1088/424 =  &  384/424 =   \\ \hline
Exact&  0.07547... & 0.45283...  & 2.56603...  & 0.90566...   \\ \hline
CS   &  0.0759(11) & 0.4543(35)  & 2.5672(43)  & 0.9026(31)   \\ \hline
RW   &  0.2216(08) & 0.5471(07)  & 2.3080(11)  & 0.9233(10)   \\ \hline
\end{tabular}
\caption{The $h_2(n)$ histogram results for the $L=4$ lattice (on this
lattice there are 424 configurations at $E/N=-1$).} 
\end{table}

\begin{description}

\item{Figure 1:} Canonical and RW histograms $h_2(n)$ at $E/N=-1$ with
$n$ given by (\ref{n}). All the canonical histograms collapse onto the
highest curve. The RW histograms follow in the order $L=10, 20, 40$
and $80$ from up to down.

\end{description}


\begin{thebibliography}{12}

\bibitem{ToVa77} G.M. Torrie and J.P. Valleau, J. Comp. Phys. {\bf 22},
                 187 (1977).

\bibitem{Be97} B.A. Berg, in:
               {\em Multiscale Phenomena and Their Simulation\/},
               Proceedings of the International Conference, Bielefeld, 
               Sept.~30 -- Oct.~4, 1996, eds. F. Karsch, B. Monien, and 
               H. Satz (World Scientific, Singapore, 1997), pp.~137-146. 

\bibitem{all} B.A. Berg, U. Hansmann, and T. Neuhaus, Z. Phys. {\bf 90}, 
 229 (1993); N.B. Wilding and M. M\"uller, J. Chem. Phys. {\bf 102}, 2562 
 (1995); W. Janke and S. Kappler, Phys. Rev. Lett. {\bf 74}, 212 (1995).
 
\bibitem{Be98} B.A. Berg, Nucl. Phys. B (Proc. Suppl.) {\bf 63A-C}, 982
               (1998); J. Stat. Phys. {\bf 82}, 323 (1996).

\bibitem{Be93} B.A. Berg, Nature {\bf 361} (1993) 708; 
               Comp. Phys. Commun. {\bf 98} (1996) 35.

\bibitem{HeOl98} P.M.C. de Oliveira, T.J.P. Penna and H.J. Herrmann, Braz.
        J. Phys. {\bf 26} (1996) 677; Europ. Phys. J. {\bf B1} (1998) 205.
After submitting our paper we were informed about one further reference:
P.M.C. de Oliveira, Int. J. Mod. Phys. {\bf C 9} (1998) 497.

\bibitem{cut} Without this cut-off the RW would extend all the way to
$E/N=+2$, where a method of return has then to be imposed anyway. In 
our RW simulations slowing down (measured in CPU time) is with the square 
(or worse) of the length of the RW. Hence, without this cut-off CPU time 
consumption would increase by at least a factor of four.

\bibitem{1d} For the $1d$ Ising model RW and microcanonical expectation 
values for $\overline{N}_i(E)$ may agree due to the trivial character of 
the model: It has only one FG for $i\ge1$ and one FG for $i\le -1$.
\end{thebibliography}
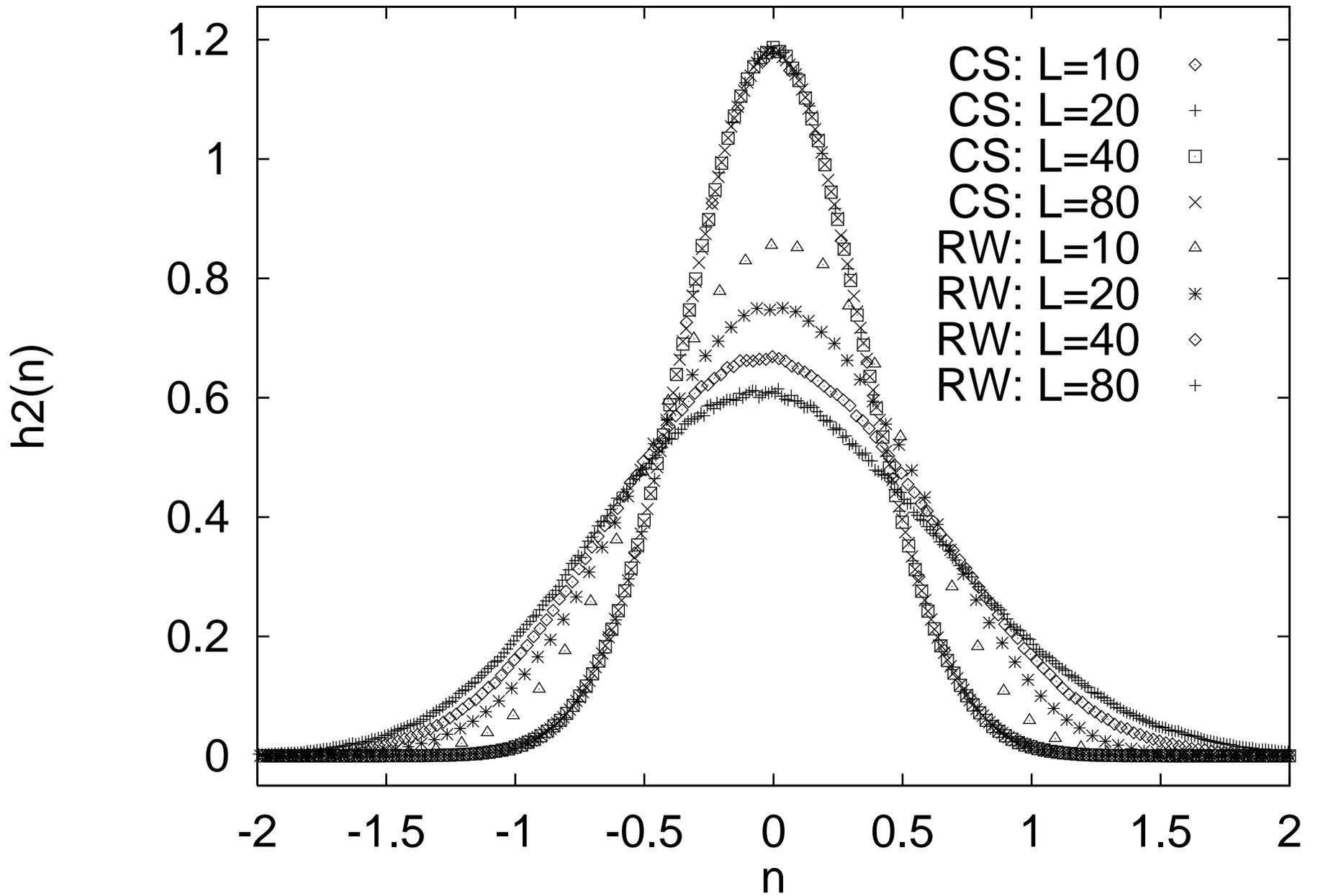
\end{document}